\PassOptionsToPackage{usenames,dvipsnames,table}{xcolor}
\documentclass{article}

\usepackage{microtype}
\usepackage{graphicx}
\usepackage{booktabs} 

\usepackage{hyperref}

\usepackage[accepted]{icml2024}

\usepackage{amsmath}
\usepackage{amssymb}
\usepackage{mathtools}
\usepackage{amsthm}
\usepackage{pifont}

\usepackage[capitalize,noabbrev]{cleveref}

\theoremstyle{plain}

\theoremstyle{definition}

\theoremstyle{remark}

\usepackage[textsize=tiny]{todonotes}

\usepackage{algorithm}
\usepackage{algpseudocode}
\usepackage{subcaption}
\usepackage[usenames,dvipsnames,table]{xcolor}

\usepackage{xspace}

\newcommand{\naive}{naïve\xspace}

\newcommand{\var}[1]{%
  \textit{#1}\xspace%
}

\newcommand{\func}[1]{%
  \textsf{\small#1}\xspace%
}

\newcommand{\NOTE}[1]{\phantom{}\begingroup\relax\ifmmode\boldmath\else\bfseries\fi\color{Cerulean}\ignorespaces#1\ignorespaces\endgroup}
\newcommand{\TODO}[1]{\phantom{}\begingroup\relax\ifmmode\else\sffamily\fi\color{BurntOrange}\ignorespaces#1\ignorespaces\endgroup}
\newcommand{\FIXME}[1]{\phantom{}\begingroup\relax\ifmmode\boldmath\else\bfseries\sffamily\fi\color{Red}\ignorespaces#1\ignorespaces\endgroup}
\newcommand{\FIXED}[1]{\phantom{}\begingroup\relax\ifmmode\else\sffamily\fi\color{Green}\ignorespaces#1\ignorespaces\endgroup}
\newcommand{\DELETE}[1]{\phantom{}\begingroup\relax\ifmmode\else\sffamily\fi\color{Red}\ifmmode\text{\sout{\ensuremath{#1}}}\else\sout{\ignorespaces#1\ignorespaces}\fi\endgroup}

\Crefname{section}{Section}{Sections}
\crefname{section}{\S}{\S}
\crefformat{section}{#2\S#1#3}

\Crefname{figure}{Figure}{Figures}
\crefname{figure}{Fig.}{Figs.}

\Crefname{equation}{Equation}{Equations}
\crefname{equation}{Eq.}{Eqs.}

\Crefname{algorithm}{Algorithm}{Algorithms}
\crefname{algorithm}{Alg.}{Algs.}

\usepackage[inline]{enumitem}

\newenvironment{myitemize}{
\begin{itemize}[nosep,wide,leftmargin=.9em]
}{\end{itemize}}

\begin{document}


\newcommand{\myeqfont}{\small}




\setlength{\textfloatsep}{10pt plus 1.0pt minus 2.0pt}
\setlength{\intextsep}{6pt plus 1.0pt minus 2.0pt}

\captionsetup[figure]{skip=2pt}
\captionsetup[table]{skip=2pt}

\setlength{\abovedisplayskip}{0pt}
\setlength{\belowdisplayskip}{0pt}
\setlength{\abovedisplayshortskip}{40pt}
\setlength{\belowdisplayshortskip}{40pt}

\clubpenalty=10000
\widowpenalty=10000
\displaywidowpenalty=10000


\twocolumn[
\icmltitle{SAM-I-Am: Semantic Boosting for Zero-shot Atomic-Scale Electron Micrograph Segmentation}



\icmlsetsymbol{equal}{*}

\begin{icmlauthorlist}
\icmlauthor{Waqwoya Abebe}{iastate,ACMD}
\icmlauthor{Jan Strube}{ACMD,uofo}
\icmlauthor{Luanzheng Guo}{ACMD}
\icmlauthor{Nathan R. Tallent}{ACMD}
\icmlauthor{Oceane Bel}{ACMD}
\icmlauthor{Steven Spurgeon}{PDD}
\icmlauthor{Christina Doty}{PDD}
\icmlauthor{Ali Jannesari}{iastate}
\end{icmlauthorlist}


\icmlaffiliation{ACMD}{Advanced Computations and Maths Division, Pacific Northwest National Laboratory, Richland, WA, USA}
\icmlaffiliation{PDD}{Physical Detection Systems and Deployment, Pacific Northwest National Laboratory, Richland, WA, USA}
\icmlaffiliation{uofo}{Institute for Fundamental Science, University of Oregon, Eugene, OR, USA}
\icmlaffiliation{iastate}{Iowa State University, Ames IA, USA}

\icmlcorrespondingauthor{Waqwoya Abebe}{wmabebe@iastate.edu}

]









\printAffiliationsAndNotice{}

\begin{abstract} 
Image segmentation is a critical enabler for tasks ranging from medical diagnostics to autonomous driving.
However, the correct segmentation semantics --- where are boundaries located? what segments are logically similar? --- change depending on the domain, such that state-of-the-art foundation models can generate meaningless and incorrect results.
Moreover, in certain domains, fine-tuning and retraining techniques are infeasible:
  obtaining labels is costly and time-consuming;
  domain images (micrographs) can be exponentially diverse; and
  data sharing  (for third-party retraining) is restricted.
To enable rapid adaptation of the best segmentation technology, we propose the concept of \emph{semantic boosting}: given a zero-shot foundation model, \emph{guide} its segmentation and adjust results to match domain expectations.
We apply semantic boosting to the Segment Anything Model (SAM) to obtain \emph{microstructure segmentation} for transmission electron microscopy.
Our booster, SAM-I-Am, extracts geometric and textural features of various intermediate masks to perform mask removal and mask merging operations.
We demonstrate a zero-shot performance increase of (absolute) +21.35\%, +12.6\%, +5.27\% in mean IoU, and a -9.91\%, -18.42\%, -4.06\% drop in mean false positive masks across images of three difficulty classes over vanilla SAM (ViT-L).
\end{abstract}












\section{Introduction}
\label{sec:introduction}

Modern materials science aims to understand and manipulate the structure of matter to obtain materials for energy conversion and storage, biomedicine, and new computing paradigms.
Transmission electron microscopy (TEM) is one of the highest resolution, most data-rich platforms to investigate the intricate details of nanoscale materials and biological specimens \cite{spurgeon2020scanning}.
TEM enables the study of matter down to the atomic level with exceptional structural and chemical sensitivity. New automated experimental paradigms can now produce TEM data at unprecedented rates, driving a need for domain-specific, adaptable, and robust machine learning-based approaches to analysis.\cite{Kalinin.2023}

Unfortunately, it has been challenging to develop domain-specific analytics that extract meaningful information from raw microscope data \cite{Kalinin2023}. A central problem in TEM analysis is the accurate and precise segmentation of microstructural image features \cite{akers2021rapid, stuckner2022microstructure}. This task is difficult to automate because of the millions of potential crystal structures, defects, and artifacts in microscopy data\cite{nextGenSpurgeon2021}. Complex interactions between the electron beam and sample~\cite{nextGenSpurgeon2021} resulting from varying imaging conditions, sample orientation, sample thickness, and detector configuration, greatly change the representation of an object in collected data. As a result, unknown images are often poorly described by (limited) prior datasets.

This data sparsity is a fundamental barrier to the use of many ML-based computer vision approaches. While low-cost synthetic images generated via approaches such as multislice simulation~\cite{Rangel2021} can be used to train models, these require extensive a priori assumptions about samples and imaging conditions. Alternatively, few-shot approaches based limited human-generated feature labels have shown some success but can fail to generalize~\cite{akers2021rapid}. 
Thus, there is a critical need for new analytics to accelerate material discovery in TEM, particularly in emerging automated microscopy that combines analytics with theory-driven hypothesis generation \cite{Liu.2022mhe}.




Here we introduce an approach that harnesses the power of a zero-shot foundation model for TEM image segmentation. Most state-of-the-art TEM-segmentation techniques use convolutional neural architectures such as U-Net \cite{ronneberger2015u}. Models are trained on microscopy data that is carefully labeled by domain experts with substantial time and cost. Moreover, it remains challenging for models trained in such manner to adapt to out-of-distribution data. 

\subsection{Related work}
\label{sec:related}

Deep learning-based image segmentation has been applied to TEM images for specific tasks such as separating atomic resolution phases, and rapid determination of microstructural features \cite{kalinin2022deep}. Many past studies utilize the U-Net architecture, e.g.,~\cite{horwath2020understanding} for the segmentation of supported gold nanoparticles. These studies discuss the effect of image resolution on segmentation accuracy, and the role of regularization and preprocessing in controlling variance. These studies further show how image features are learned, so that model architectures can be better designed depending on the task at hand. \cite{groschner2021machine} combines a U-Net image segmentation model with random forest classifier for detecting stacking faults in TEM images of Au and CdSe nanoparticles. \cite{sadre2021deep} compares the performance of conventional Bragg filtering 
with a U-Net for phase-contrast images of monolayer graphene which can produce complex nonlinear contrast.

\cite{bell2022trainable} proposes trainable segmentation for TEM images. The authors start by manually labeling a few background and particle pixels. The labelled pixels and corresponding feature values are used to train a classifier. The classifier is used to differentiate boundaries between labelled sets of particle and background pixels. This process yields a fully masked image and a trained classifier which is used to classify more images. 
The process relies on less manual labeling than training CNN-based segmentation models. \cite{kim2020unsupervised} presents an unsupervised learning based on convolutional neural networks and a superpixel algorithm for segmentation of a low-carbon steel microstructure.

\cite{akers2021rapid} proposes a semi-supervised few-shot machine learning approach for TEM segmentation of three oxide material systems, (1) SrTiO$_3$ / Ge epitaxial heterostructures, (2) La$_{0.8}$Sr$_{0.2}$FeO$_3$ thin films, and (3) MoO$_3$ nanoparticles. Given a fine-tuning set of less than 10 tiny crops (chips) of the image that represent micro-structural features, their model is able to make mask predictions comparable to those of domain experts.  However, such masks are limited to chip-level representations. \cite{stuckner2022microstructure}, proposes transfer learning from MicroNet (a large dataset of microscopy images) for improving the mean intersection over union (mIoU) of microscopy segmentation tasks. They show that models pre-trained on their dataset exhibited superior generalization to new microscopy images compared to models trained on other datasets like ImageNet.

AtomAI \cite{ziatdinov2021atomai} produced an open-source software package combining instrument-specific Python libraries, deep learning, and simulation tools. The authors implement both U-Net and customized CNN segmentation models. The semantic segmentation task categorizes each pixel in the input image as belonging to a particular object, such as atom type, defect structure, or background. The authors provide convenient APIs for instantiating and training their various models. TEMImageNet~\cite{lin2021temimagenet} implemented a training library and deep learning models for atom segmentation, localization, denoising, and super-resolution processing of experimental images. Although 
trained on simulated data, the author's models 
generalize to microscopy data. \cite{khan2023leveraging} proposes using adversarial networks to rapidly generate large microscopy datasets for training downstream tasks such as TEM segmentation.

Since these studies generally define their goal as a semantic segmentation task, they train models to map pixels to specific pixel-level labels. Consequently, their models entirely rely on labeled TEM images for training and are ultimately limited to segmenting just a small subset of materials. i.e. they are unable to deal with out-of-distribution data. In contrast, we define our task, \textit{microstructure segmentation}, as a promptable segmentation task allowing us to harness the potential of SAM. This minimizes our reliance on labeled TEM data, which is expensive to produce, and provides the versatility to segment out-of-distribution data. 












\subsection{Overview}

In this work, we consider the common case of cross-sectional images of thin film interfaces, which consist of atomically-resolved crystalline (periodic) structures projected onto a 2-D surface~\cite{Spurgeon.20177bs}.
These images typically consist of two or more distinct chemical compounds, as well as defects, defined as deviations from the ideal lattice structure.
Segmentation of such images is  challenging because of the low contrast between the nanomaterials and substrate \cite{groschner2021machine}. This difficulty is exacerbated by the various irregular shapes and the presence of certain materials defects (e.g. antiphase boundaries, grain boundaries, dislocations, and voids).
Moreover, the lack of ground truth labels only increases the challenge.
Finally, often there are restrictions on sharing TEM data that prevent accumulation of larger shared pools of labels.
The restrictions also prevent sharing data to enable third-party retraining or fine-tuning of models.

\begin{figure}[t]
    \centering
    \includegraphics[width=\columnwidth]{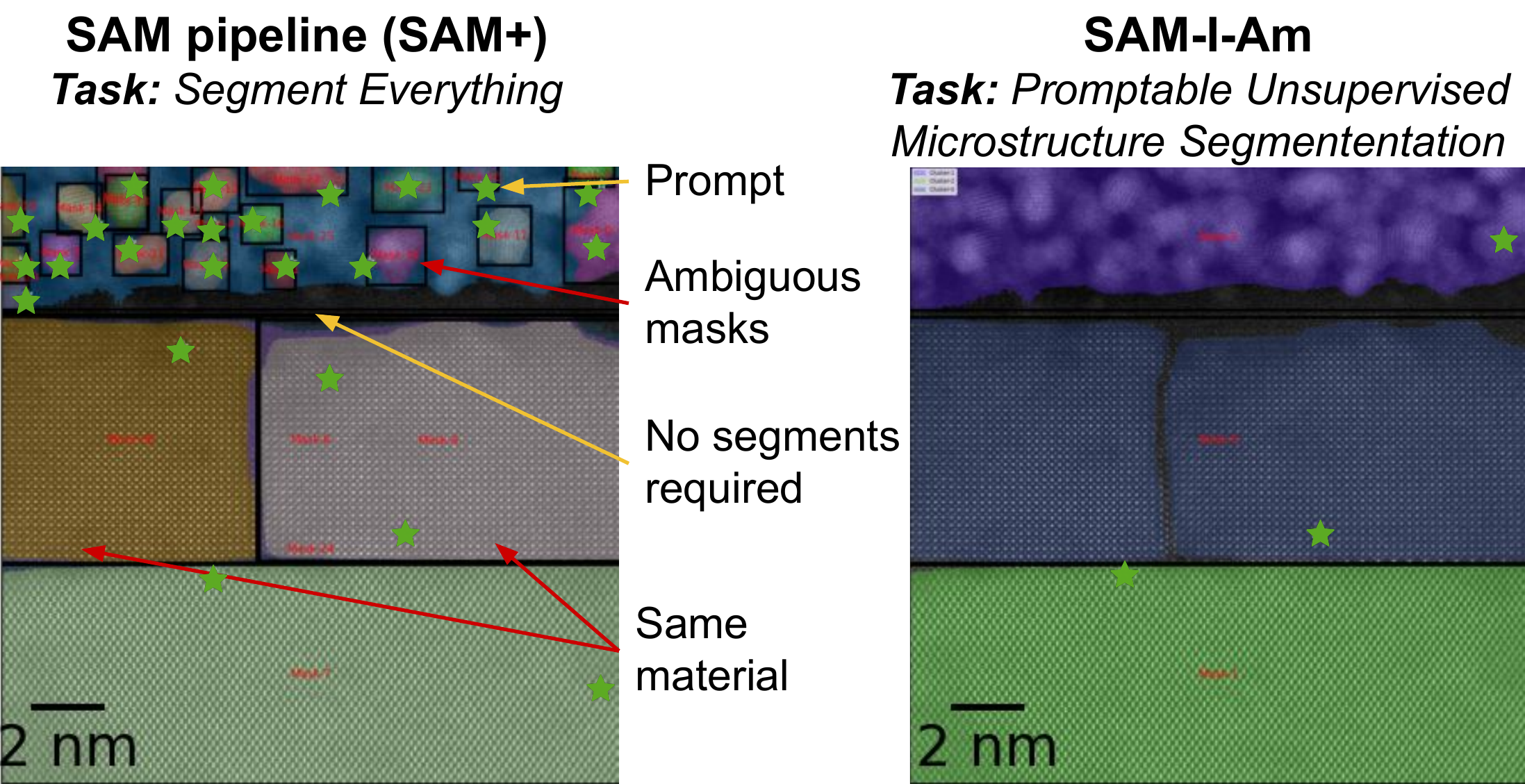}
    \caption{Comparing the vanilla SAM pipeline and SAM-I-Am for a cross-sectional image of three layers: Pt / C (top), SrTiO$_3$ (middle) and Ge (bottom). The SAM pipeline conducts the `segment anything' task that yields ambiguous masks and fails to identify surfaces with similar material makeup. Using semantic boosting, SAM-I-Am delivers the \textit{microstructural segmentation} task.}
    \label{fig:comparison_1}
\end{figure}

To address these challenges, we propose a new task called \emph{microstructure segmentation} and demonstrate how to achieve it with SAM-I-Am%
\footnote{
  Our name honors the trio of the famous character in Dr.\@ Seuss' semantically amusing \emph{Green Eggs And Ham}, SAM's achievement, and 
  \textbf{I}dentification of \textbf{A}tomic-scale \textbf{m}icrographs. \href{https://github.com/PerfLab-EXaCT/SamIAm/graphs/contributors}{Github}. 
}%
, a lightweight zero-shot \emph{semantic booster} for the Segment Anything Model (SAM) \cite{kirillov2023segment}.
Given an input micrograph and prompt, \textit{microstructure segmentation} requires microstructures belonging to the same materials to be segmented under the same masks.
SAM-I-Am converts \naive mask outputs of the SAM pipeline into microstructure masks.
It relies on SAM to recognize unseen materials without retraining and fine-tuning, and is thus zero-shot.
It is \emph{lightweight} in that its internal computations are significantly cheaper compared to the SAM pipeline it augments. 

To the best of our knowledge, the SAM-I-Am semantic booster is the first zero-shot promptable foundation model for the TEM domain.


Although SAM's segment everything task \cite{zhang2023faster} resembles ours, \emph{microstructure segmentation} has fundamental differences, illustrated in Fig.~\ref{fig:comparison_1}.
The figure shows three distinct materials stacked horizontally.
SAM's segmentation is on the left.
In the top layer, SAM identifies materials in another (atomically non-composite) material, which is nonsense and is analogous to LLM hallucinations.
In the middle layer, SAM incorrectly identifies the two halves as distinct materials.
Finally, the dark area under the top region is gap and requires no segment.
The cumulative effect of these errors means the results require human intervention to avoid \emph{invalid} hypotheses that \emph{misdirect} future experiments.

SAM-I-Am's semantic boosting produces the correct microstructure segmentation on the right of Fig.~\ref{fig:comparison_1}.
Our contributions are as follows: 
\begin{myitemize}

\item Define \textit{microstructure segmentation}, a promptable unsupervised semantic segmentation task and distinguish from `semantic segmentation'.


  \item Propose a lightweight method of \emph{semantic boosting} for microstructure segmentation.
  We define \emph{microstructure segmentation} as a promptable task and leverage a foundation segmentation model.
  We minimize reliance on expensive labels and avoid sharing of sensitive data.


  \item Demonstrate zero-shot performance and achieve +21.35\%, +12.6\%, +5.27\% in mean IoU, and a -9.91\%, -18.42\%, -4.06\% drop in mean false positive masks  across images of three difficulty classes over vanilla SAM (ViT-L) pipeline.
    
  \item Propose both unsupervised and an alternative supervised semantic boosting, implemented in the open-source SAM-I-Am tool.


\end{myitemize}
Semantic boosting opens new horizons in TEM research, offering busy researchers the tools to unlock the full potential for high-resolution TEM image segmentation.
Our work also contributes to the broader field of computer vision and image analysis, showcasing the adaptability and power of advanced zero-shot foundation models in challenging and specialized domains.




\section{Results and Discussion}
\label{sec:overview}







\begin{figure*}
    \centering
    \includegraphics[width=0.9\textwidth]{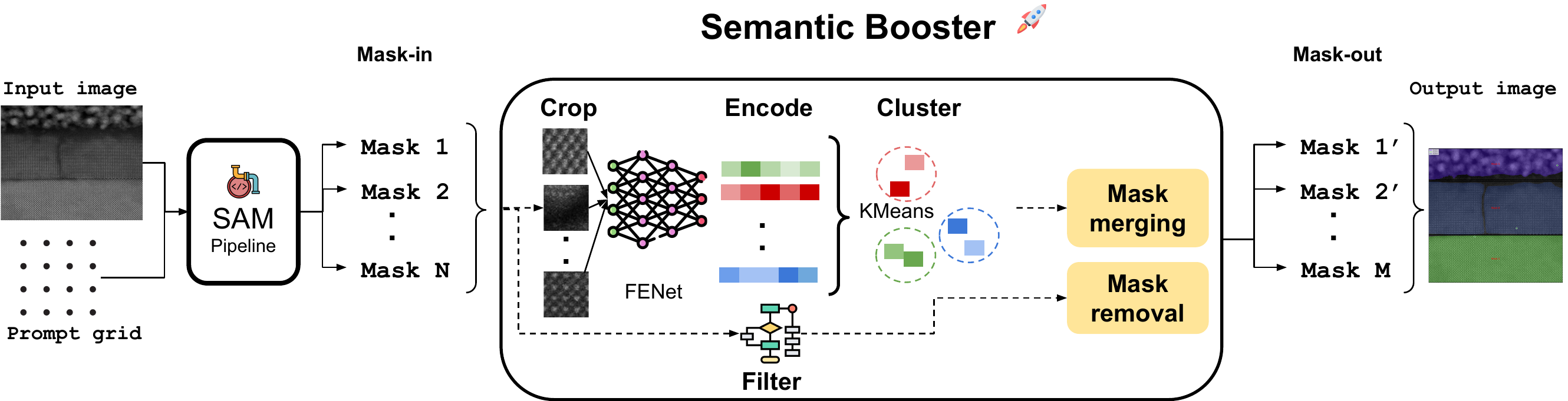}
    \caption{Semantic boosting: The proposed booster augments SAM+ by performing a mask-in mask-out post-processing procedure for mask removal and mask merging operations.}
    \label{fig:pipeline}
\end{figure*}

SAM provides an `AutoMaskGenerator' pipeline that we refer to as SAM+.
SAM+ autoprompts the SAM model to simultaneously produce mask predictions (segmentations) for all objects in an input image.
SAM+ is used for the `segment everything' task, which resembles our \textit{microstructure segmentation} task. It takes a set of parameters to compute the resulting masks. More importantly, it performs pre- and post-processing such as layering, cropping, encoding and mask de-duplication to capture all artifacts in the input image.  Our proposed approach (Fig.~\ref{fig:pipeline}) 
takes SAM+'s outputs and applies semantic boosting---filtering, correcting, and enhancing---to generate material microstructural segments.

\subsection{Limitations of SAM+}





SAM+ provides a mechanism to specify parameters that control autoprompts.
Behind the scenes, it dissects an input image into layers where each layer contains equal sized crops of the image obtained by applying a sliding crop window. Next, each crop belonging to layer$_i$ is prompted by projecting (overlaying) an $n_i \times n_i$ grid of points to identify potential regions. A typical parameter, like the `prompt grid length', for instance, specifies the size of input prompt for the first layer ($n_0$). Given prompt density for the first layer's crop, crops of subsequent layers will be prompted with sparser points. i.e. $n_{i+1} < n_i$.

Unfortunately, SAM+'s vanilla parameters are inadequate for achieving \textit{microstructure segmentation}.
Consider Fig.~\ref{fig:comparison_1}, which shows a sample with three distinct materials stacked horizontally.
SAM+'s segmentation is on the left.
For the middle layer (SrTiO$_3$), SAM+ generates two separate masks. However, the whole middle section is actually the same material `split' by an antiphase boundary defect (typically of special interest). This shows SAM has no semantic awareness to combine these masks.
The narrow gap between the top and middle needs no segment. (Here, SAM+ correctly avoids segmenting it.)
For the top layer (Pt / C), SAM generates masks for several subregions inside a larger region.
The result implies that a single (atomically non-composite) material may contain a second material, which is semantic nonsense.
This behavior is called `ambiguity'. Ambiguity arises when a SAM prompt causes uncertainty about how to treat a region and its immediate surrounding (parent region). In response, SAM generates masks for both the region and its parent.

A fundamental problem is that the \emph{segment everything} task fails to properly address the \emph{microstructure segmentaion} problem. Therefore, it is imperative to define the task in order to improve the tool.

\subsection{Problem statement}

In contrast to semantic segmentation, the goal of \textit{microstructure segmentation} is not to map every pixel to pre-defined classes (labels).  Instead, our focus lies on a high-level perspective of identifying and delineating surfaces of arbitrary materials. Moreover, unlike semantic segmentation, all pixels of an image may or may not be segmented. 

\textit{Microstructure segmentation} also differs from the `segment everything' task because it does not permit mask `ambiguity'. Consider the case of prompting an image of a car by placing a prompt on the car door. SAM might respond to this ambiguous prompt with 2 overlapping masks, one for the car door and another for the entire body of the car. Both are considered valid under the `segment everything' task. On the contrary, in our task, overlapping masks are nonsensical. A sub-region and a surrounding parent region are either the same material (requiring combination), or different materials (requiring dissolution). In the latter case, the parent mask must not completely engulf the child mask, (i.e. it should simply encircle it instead).

At a high-level, we define \textit{microstructure segmentation} as a promptable unsupervised semantic segmentation task, a more abstract usecase with three goals:
%
%
\begin{myitemize}
\item \textbf{For each material, mask the entire region covered by the material}: Here, pixels belonging to the same material must also belong to the same mask. i.e. $\var{pixel}_{ij}  \in \var{material}_x \; \Leftrightarrow \var{pixel}_{ij}  \in \var{mask}_x$.

  \item  \textbf{Resolve ambiguous masks:} Resolve ambiguous (false positive) masks that cover semantically meaningless regions in the input image. Ideally, we desire a one-to-one mapping \( f \) from predicted masks \( M \) to ground truth segments \( S \).

  \item \textbf{Perform without any specific knowledge of materials:}
    At least one masking mode must assume no knowledge of materials other than universal properties.
    We permit other modes with \emph{supplemental} customization, but this work does \emph{not} employ it. 
\end{myitemize}

Formally, given a set $S$ containing subareas of an image $I \in \mathbb{R}^{a \times b}$, prompt $P = \{(x, y) \mid x, y \in \mathbb{Z^+}\}$, and objective function $f$, the task is to generate a set $M$ that closely matches $S$ in terms of  cardinality and overlap of corresponding subsets, i.e. $f : I \times P \rightarrow M \approx S$. This calls for both minimizing the cardinality of $M$ to ideally match that of $S$, while maximizing the overlap between corresponding pairs in $M$ and $S$. Let:


\begin{myitemize}
    \item $S$ : be the ground truth set corresponding to $I$.
    \item $M$ : be the set of masks generated by $f$ for $(I, P)$.
    \item $|M|$ : be the cardinality of set $M$.
    \item $J(S_i,M_i)$ : be the Jaccard index between $S_i$ and $M_i$.
\end{myitemize}

\textbf{Objective function 1:}
\begin{equation}
\begin{aligned}
& \text{Maximize } \sum_{i} J(S_i, M_i) - \lambda |M| \\
& \text{Subject to } \forall i, M_i \sim S_i
\end{aligned}
\end{equation}

where, the $\lambda$ parameter controls the trade-off between maximizing overlap and minimizing cardinality.


\subsection{Semantic boosting}

\begin{algorithm}[t]
\caption{Semantic Booster}
\label{alg:postprocess}
\begin{algorithmic}[1]
\State expendables $\gets$ []

\For{mask in masks}
    \State children, redundant $\gets$ compute\_related(mask)
    
    \For{child in children}
        \If{child.area $<$ 5\% of mask area}
            \State expendables.add(child)
        \EndIf
    \EndFor
    
    \If{sum(children area) $>$ 70\% of mask area}
        \State expendables.add(mask)
    \EndIf
    
    \State max\_duplicate $\gets$ max(sort([mask, redundant]))
    
    \For{duplicate in [mask, redundant]}
        \If{duplicate $\neq$ max\_duplicate}
            \State expendables.add(duplicate)
        \EndIf
    \EndFor
\EndFor

\For{mask in expendables}
    \State remove(mask)
\EndFor
\State model $\gets$ pre-trained FENet
\State labels $\gets$ CompLabels(masks, model, \#\_materials)
\State map $\gets$ dictionary(labels, masks)

\For{label, masks in map}
    \State merge(masks)
\EndFor
\end{algorithmic}%
\end{algorithm}


To enable microstructure segmentation, we apply a semantic booster to SAM+'s results.
The booster is based on the axiom that a segment is defined by its atomic microstructure.
There are several implications:

\begin{enumerate*}

\item A segment cannot be composite: masks may not overlap. 

\item The interfaces that separate different microstructures can vary: permit distinct segments to be separated by gaps.

\item Masks with the same atomic structure are the same microstructure (noting that different compounds may have highly similar 2D projections in electron micrographs requiring additional spectral data for separation).

\end{enumerate*}

The booster, shown in Fig.~\ref{fig:pipeline}, is implemented as a post-processing engine that transforms \naive masks into microstructure masks. Such resolution of masks involves mask removal and mask merging operations.
First, mask removal filters out ambiguous masks based on geometric properties. In particular, the post-processor uses geometric thresholds to filter out tiny child masks, tiny island masks, compound masks and redundant masks as shown in Algo.~\ref{alg:postprocess}. Next, mask merging utilizes a texture-based semantic model coupled with unsupervised KMeans clustering to determine the similarity between remaining masks. Similar masks are merged to span regions that cover the same microstructure. 

\subsubsection{Mask removal}

Ambiguous masks come in all shapes and sizes. For instance, some ambiguous child masks are too small in comparison to their parent, whereas in other cases, ambiguous child masks could be almost large enough to constitute the parent mask. In the first case, the filter prioritizes the parent (filters tiny children for removal), while in the latter case it prioritizes the child masks and relies on the mask merging step to resolve the children.

Additionally, certain masks almost entirely overlap with each other, indicating that they evaded the default mask de-duplication procedure. Finally, some tiny masks appear to be an island contained within a greater surrounding region that was not segmented. Algo.~\ref{alg:postprocess}, shows the mask filtering process conducted by iterating over the outputs of SAM+.

\subsubsection{Mask merging}

Although SAM+ performs mask de-duplication, it only considers the predicted mask quality (in terms of IoU), and degree of geometric overlap between masks rather than the semantic similarity between them. The mask merging operation acts as a supplementary mask de-duplicator by merging masks with similar semantic features (underlying textures). First, the procedure randomly extracts $K$-crops (of size $m \times m$) from each mask. Next, it utilizes a texture model to derive textural features by embedding the input crops into latent space vectors. Finally, a KMeans model is used to cluster the latent space vectors into $M$ categories which are known a priori. To determine which cluster a mask belongs to, a majority vote is conducted over its $K$ embeddings.

\begin{algorithm}[t]
\caption{Compute mask labels}
\label{alg:compute-labels}
\begin{algorithmic}[1]
\Function{CompLabels}{$\text{masks, model, \#\_materials}$}
    \State $\text{map} \gets \{\}$
    \State $\text{all\_embeddings} \gets []$
    \For{$\text{mask}$ \textbf{in} $\text{masks}$}
        \State $\text{map[mask]} \gets []$
        \State $\text{crops} \gets \text{crop}(\text{mask})$
        \State $\text{embeddings} \gets []$
        \For{$\text{crop}$ \textbf{in} $\text{crops}$}
            \State $\text{embeddings.append}(\text{model}(\text{crop}))$
            \State $\text{all\_embeddings.append}(\text{model}(\text{crop}))$
        \EndFor
        \State $\text{map[mask]} \gets \text{embeddings}$
    \EndFor
    
    \State $\text{clusters} \gets \text{KMeans}(\text{all\_embeddings}, \text{\#\_materials})$
    \State $\text{labels} \gets []$
    
    \For{$\text{mask}_i, \text{embeddings}$ \textbf{in} $\text{map}$}
        \State $\text{labels.append}(\text{max(clusters(embeddings)))}$

    \EndFor

    \State \Return $\text{labels}$
\EndFunction
\end{algorithmic}
\end{algorithm}

A vital aspect of this process involves choosing a robust model for accurately extracting domain semantics (i.e. textural features). Doing so determines the quality of the latent space vectors and thus the clustering results. In particular, the textural model has contrastive objectives: minimize the embedding distance between similar crops, and maximize it for differing crops. Formally, let: 

\begin{myitemize}
    \item $\var{crop}_i,\var{crop}_j$ : be crops extracted from materials.
    \item $\var{emb}_i, \var{emb}_j$ : be the latent space vectors of the crops.
    \item $d$ : be the Euclidean distance between vectors.
\end{myitemize}

\textbf{Objective function 2:}
\begin{equation}
\begin{aligned}
& \text{if } \{\var{crop}_i,\var{crop}_j\} \in \var{material}_x: \\
& \quad \quad \text{Minimize } d(\var{emb}_i, \var{emb}_j)  \\
& \text{else:} \\
& \quad \quad \text{Maximize } d(\var{emb}_i, \var{emb}_j)
\end{aligned}
\end{equation}

Moreover, in line with our zero-shot requirements, we want to use a pre-trained model and rely on its transfer learning capability. Therefore, selecting both a robust model architecture and a pre-training task is critical.


\section{Evaluation}
\label{sec:evaluation}


\subsection{Data and model}

A primary challenge in evaluating the proposed method was the lack of labeled TEM images. Accordingly, we resorted to preparing our own validation dataset using microscopy images and labeling them using an annotation tool. To accelerate labeling enough to convince busy scientists to participate, we utilized Label-Studio \cite{LabelStudio}, an open-source image annotation tool, alongside a SAM ML-backend  (also open-source) developed by HumanSignal for AI-assisted annotation. We made some modifications to the ML backend to add support for all the SAM ViT versions. SAM-assisted labeling accelerated our labeling effort (similar to the process by which parts of the SA-1B dataset was annotated \cite{kirillov2023segment}). Before using this tool we could only obtain 3 labeled images in several hours; with it, we obtained 50 labeled images in under 2 hours.



\begin{figure}[t]
    \centering
    \includegraphics[width=.8\linewidth]{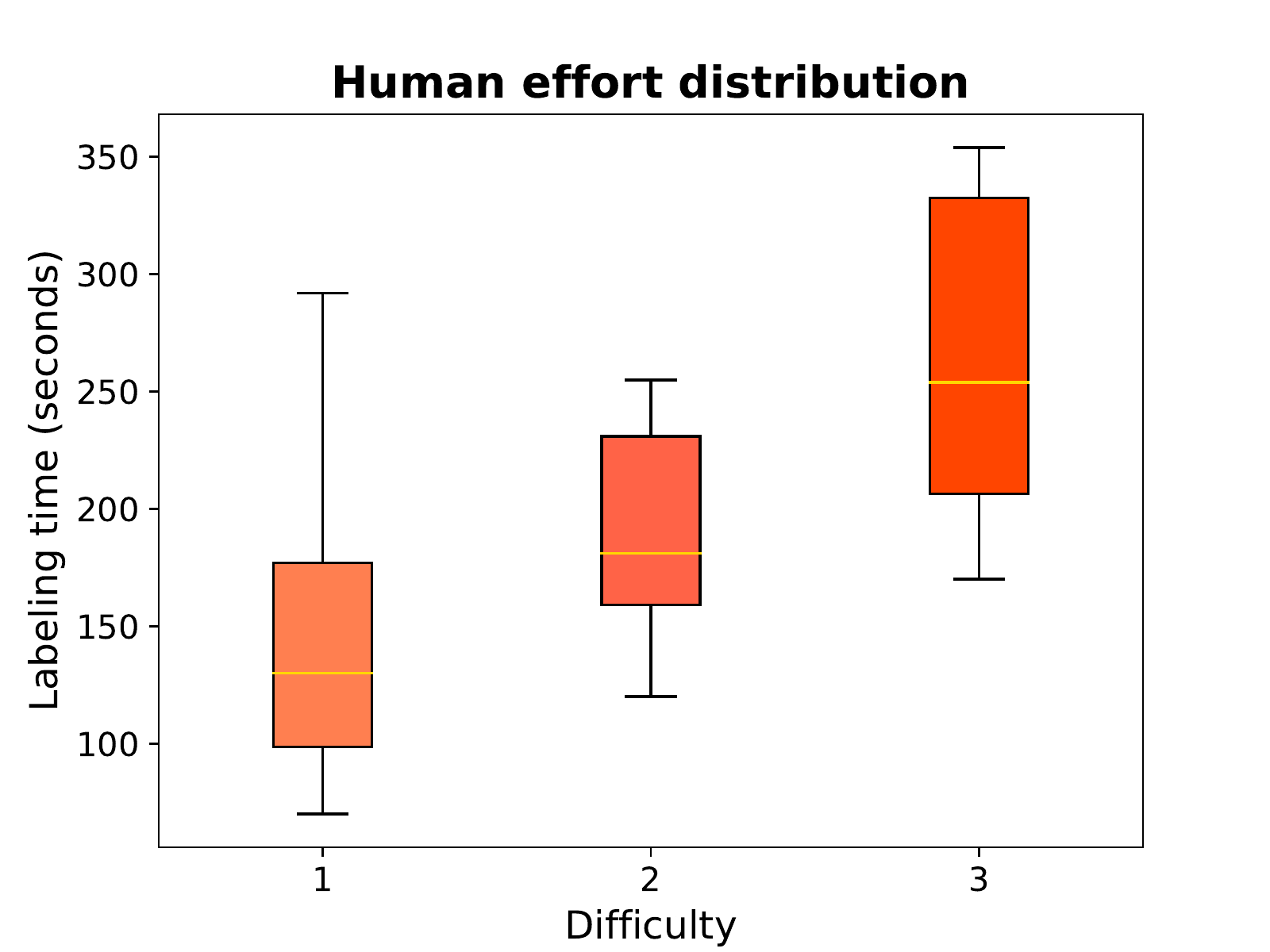}
    \caption{Labeling times in seconds for different difficulty classes of images.}
    \label{fig:effort}
\end{figure}

We acquired labels for 82 scanning TEM images, (mostly 1024 $\times$ 1024) but some with (2048 $\times$ 2048) pixel resolution, from two domain experts \cite{spurgeon_2024_10909552}. Each image contains 2 - 4 materials from among compounds including La:SrTiO$_3$,
SrTiO$_3$, Ge, Nb:SrTiO$_3$, WO$_3$, LaFeO$_3$, LaMnO$_3$, La$_{0.8}$Sr$_{0.2}$FeO$_3$, and Pt / C captured in annular dark-field (ADF) and high-angle annular dark-field (HAADF) modes. Moreover, some images show vacuum regions while some material surfaces show certain defects. As shown in Fig.~\ref{fig:effort}, the experts divided the images into 3 difficulty classes based on the intricacy of the texture patterns.

\begin{figure}[t]
    \centering
    \includegraphics[width=\columnwidth]{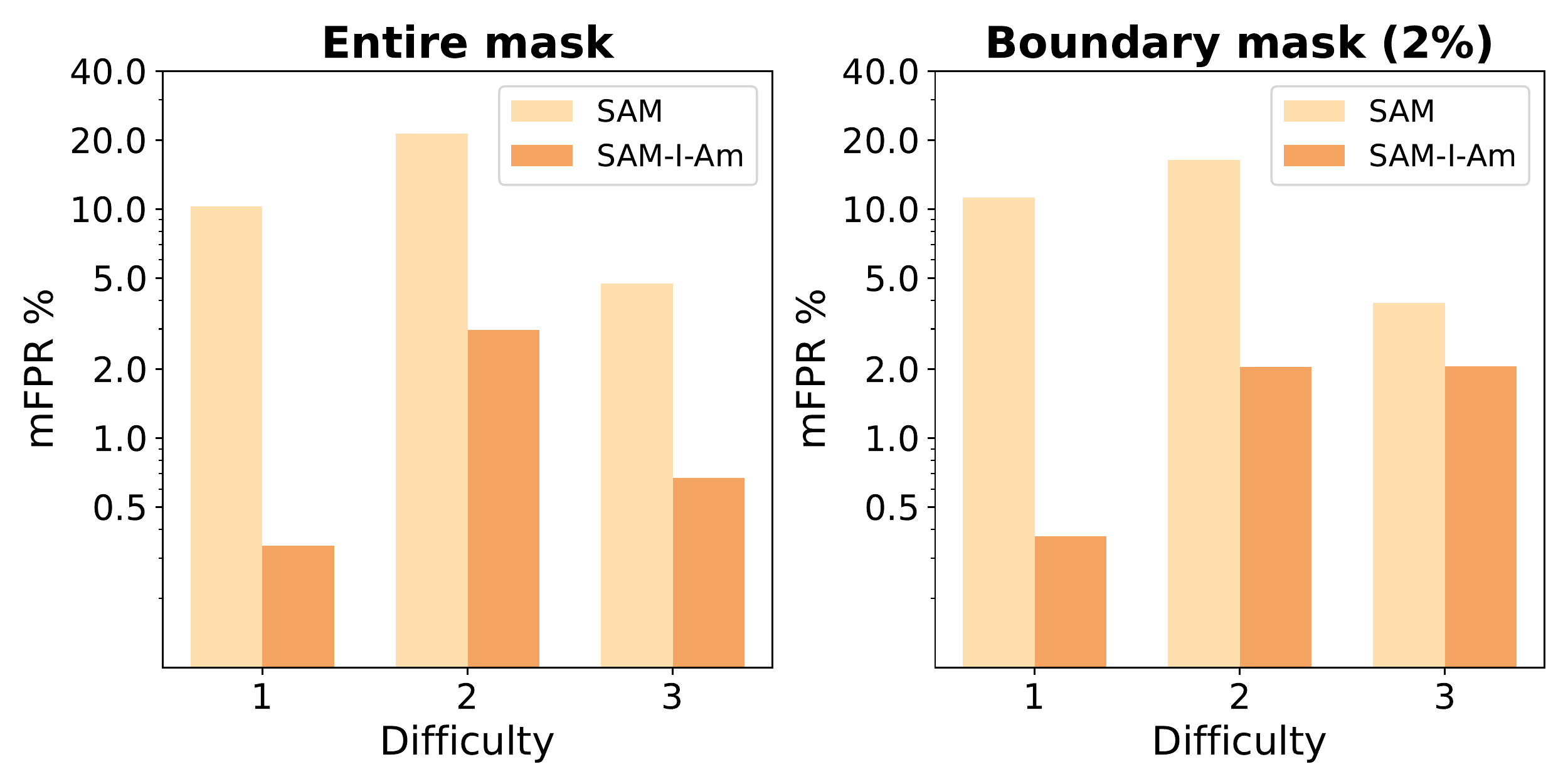}
    \caption{Comparing mean false positive rate on entire mask regions and mask boundary regions. In both cases, SAM-I-Am pipeline outperforms the SAM+ baseline.}
    \label{fig:error}
\end{figure}

The SAM ViT image encoders come in three versions, ViT-B with 91M, ViT-L with 308M, and ViT-H with 636M parameters. We run our performance (Fig.~\ref{fig:performance}) and error (Fig.~\ref{fig:error}) experiments using the ViT-L version. We applied a prompt grid size of 16 $\times$ 16. Given the nature of microscopy images we are dealing with, we observed model architectures specifically designed to detect textures performed better at encoding mask crops compared to regular convolutional backbones of comparable size. In particular, the FENet model \cite{xu2021encoding} fine-tuned on the DTD texture dataset \cite{cimpoi14describing} yielded robust crop encodings compared to a ResNet-18 model \cite{he2016deep} pre-trained on ImageNet \cite{deng2009imagenet}. We deployed an FENet with a ResNet-18 backbone and removed the last two fully connected layers for feature extraction. We used a crop size of 60 $\times$ 60 for the crop encoding step.  

\vspace{-1em}

\subsection{Metrics}

\begin{figure}[t]
    \centering
    \includegraphics[width=1\linewidth]{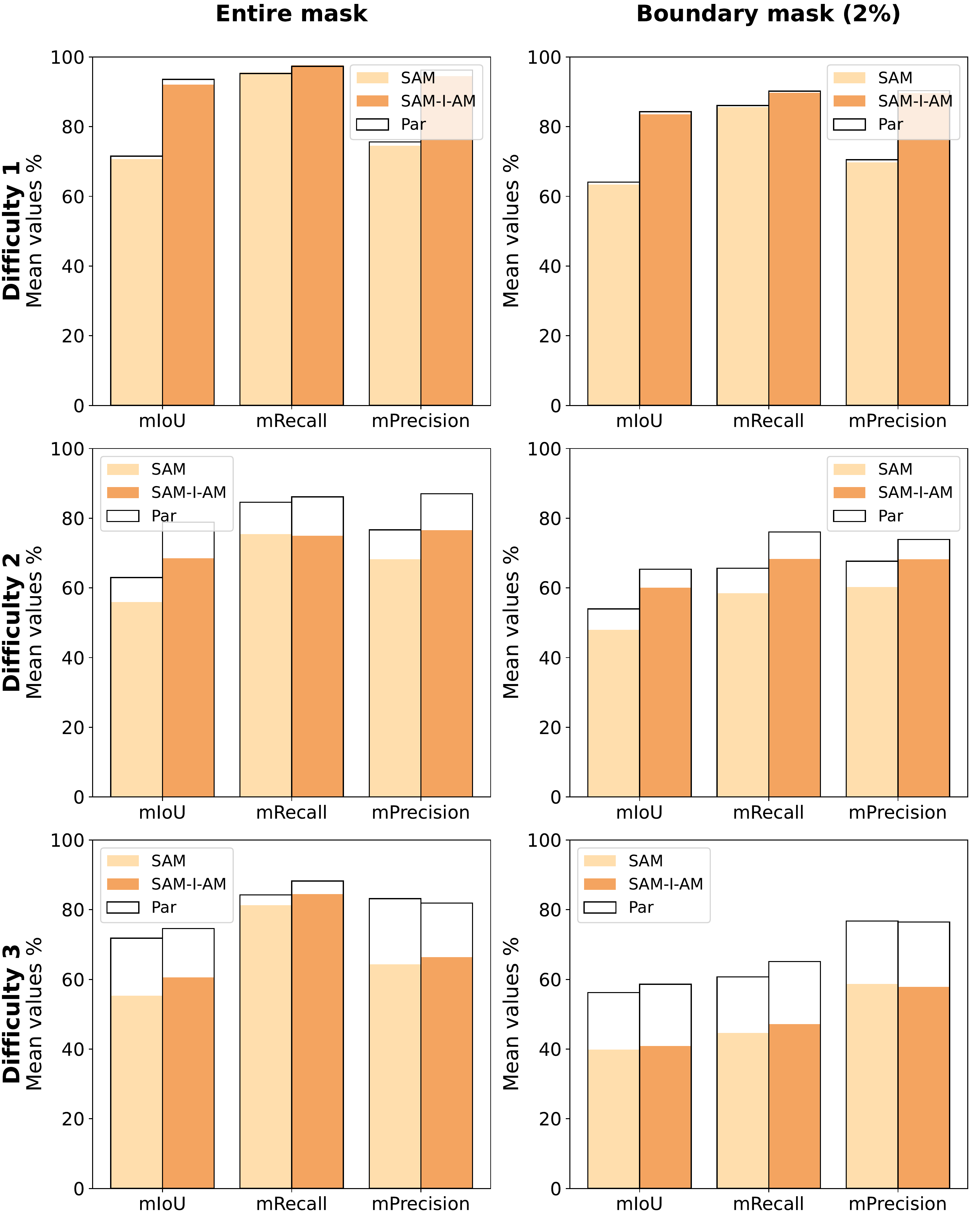}
    \caption{Comparing mean performance on entire mask regions and mask boundary regions. In both cases, SAM-I-Am pipeline outperforms the SAM+ baseline.}
    \label{fig:performance}
\end{figure}

  We report both performance and error measures using the mean IoU, mean recall, mean precision, and mean false positive rate. These scores are averaged over all images in the validation dataset. We define the metrics as follows:
\begin{equation}
    \func{mIoU} = \frac{1}{n} \sum_{k=0}^{n} \frac{1}{l}  \sum_{j=0}^{l} \frac{1}{m} \sum_{i=0}^{m} \frac {\var{mask}_{i} \bigcap \var{GT}_j}{\var{mask}_{i} \bigcup \var{GT}_j}
    \label{eqn:iou}
\end{equation}

\begin{equation}
     \func{mRecall} = \frac{1}{n} \sum_{k=0}^{n} \frac{\text{True Positive}}{\text{True Positive} + \text{False Negative}}
     \label{eqn:recall}
\end{equation}

\begin{equation}
     \func{mPrecision} = \frac{1}{n} \sum_{k=0}^{n} \frac{\text{True Positive}} {\text{True Positive} + \text{False Positive}}
     \label{eqn:precision}
\end{equation}


Here, $GT$ is the ground truth mask, $l$ is the number of ground truth labels for a given image, and $n$ specifies the number of images in the validation dataset. 

To compute the recall and precision for image $k$, we consider True Positive as the area covered by intersection between ground truth labels and masks. i.e. $\text{True Positive(k)} = \frac{1}{l} \sum_{j=0}^{l} \frac{1}{m} \sum_{i=0}^{m} (\var{mask}_i \bigcap \var{GT}_j)$. Conversely, the false positive encompasses an area covered by the predicted masks but not the ground truth. i.e. $ \text{False Positive(k)} =   \frac{1}{l} \sum_{j=0}^{l}  \frac{1}{m} \sum_{i=0}^{m} (\var{mask}_i \setminus \var{GT}_j)$. The false negative region represents an area of a ground truth mask that is not covered by the predicted mask. i.e.  $ \text{False Negative(k)} \\ = \frac{1}{l} \sum_{j=0}^{l} \frac{1}{m} \sum_{i=0}^{m} (\var{GT}_j \setminus \var{mask}_i)$. In all cases, we consider involving mask $i$ for mean metric computation with  ground truth $j$ if $\func{iou}(\var{mask}_i, \var{GT}_j) \geq 5\%$. 

\begin{figure}[t]
    \centering
    \includegraphics[width=0.4\textwidth]{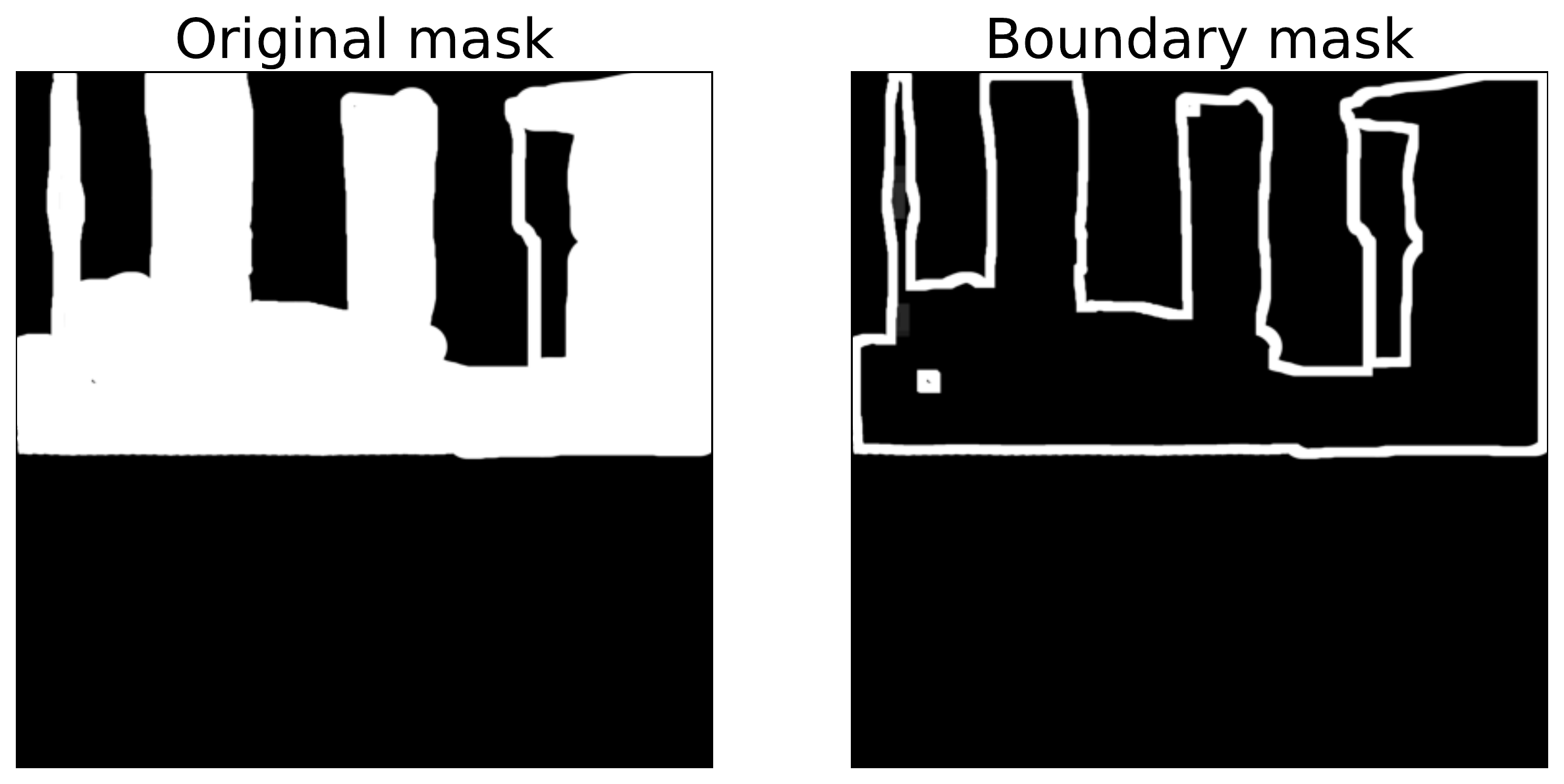}
    \caption{An original ground truth mask (left) and its corresponding boundary mask (right).
    }
    \label{fig:boundary}
\end{figure}

In order to compute the model's error measure, we implement a custom false positive rate (FPR) tailored to our downstream task. Given a ground truth mask area, we consider all non-best matching mask predictions that fall in its bounds as false positives. Hence, we compute the false positive rate as follows:
\begin{equation}\myeqfont
    \func{mFPR} = \frac{1}{n} \sum_{k=0}^{n} \frac{1}{l} \sum_{j=0}^{l} \frac{1}{m} \sum_{i=0}^{m} \func{iou}(\var{non-best mask}_i, \var{GT}_j)
    \label{eqn:fpr}
\end{equation}

We also adopt a boundaryIoU metric \cite{cheng2021boundary} to measure the overlap around the peripheries of a material. Boundary metrics are computed by first converting masks (both ground truths and predictions) into boundary masks as shown in Fig.~\ref{fig:boundary}. The boundary thickness, 2\% in our case, is measured as a percentage of the diagonal of the image. We computed all metrics for both entire mask regions as well as boundary masks as shown in Fig.~\ref{fig:performance}.

\subsection{Results}

As shown in Fig.~\ref{fig:performance}, SAM-I-Am beats SAM+ in all three difficulty classes. SAM-I-Am also has finer boundary performance in all cases. Moreover, in Fig.~\ref{fig:error}, SAM+ overproduces masks causing much higher false positive rates, especially for difficulty levels 1 and 2. In contrast, SAM-I-Am gives significantly fewer superfluous masks in all cases. 

\begin{figure}[t]
  \centering
  \begin{subfigure}{0.8\linewidth}
    \centering
    \includegraphics[width=\linewidth]{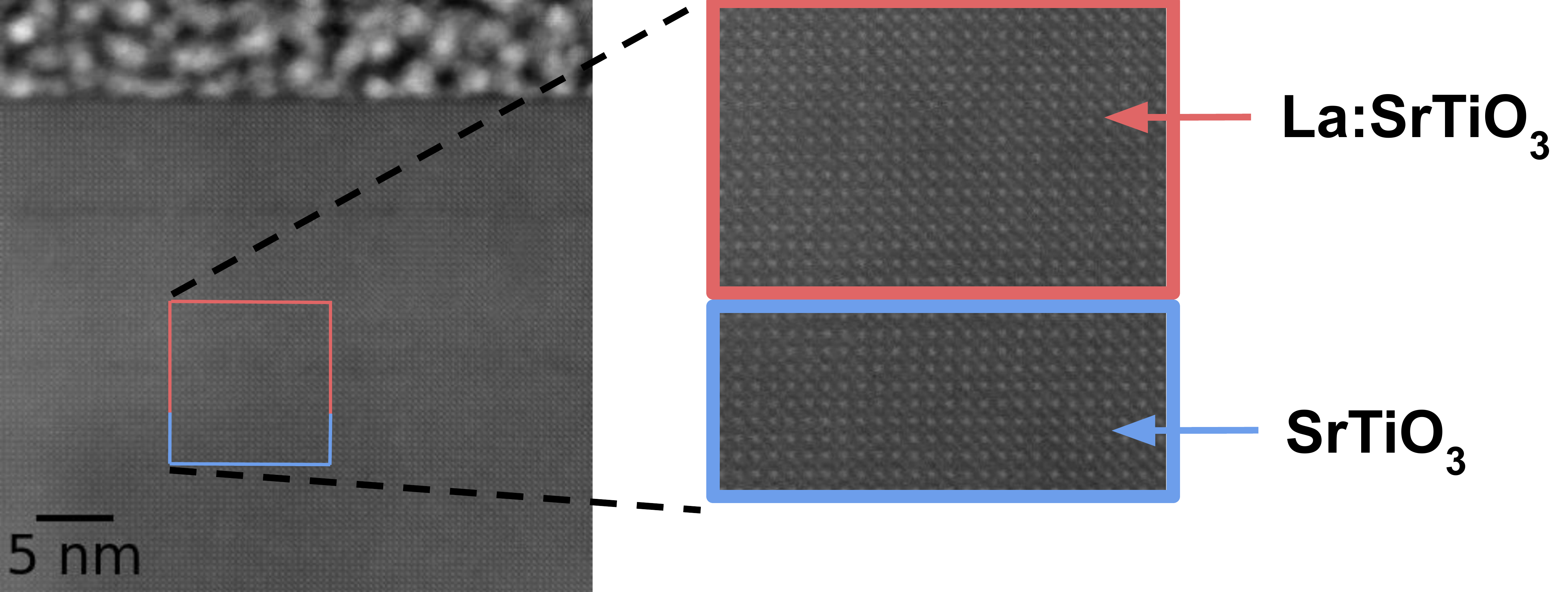}
    \caption{La:SrTiO$_3$ and SrTiO$_3$ crystals appear indistinguishable to the untrained eye.}
    \label{fig:diff_a}
  \end{subfigure}
  \begin{subfigure}{0.8\linewidth}
    \centering
    \includegraphics[width=\linewidth]{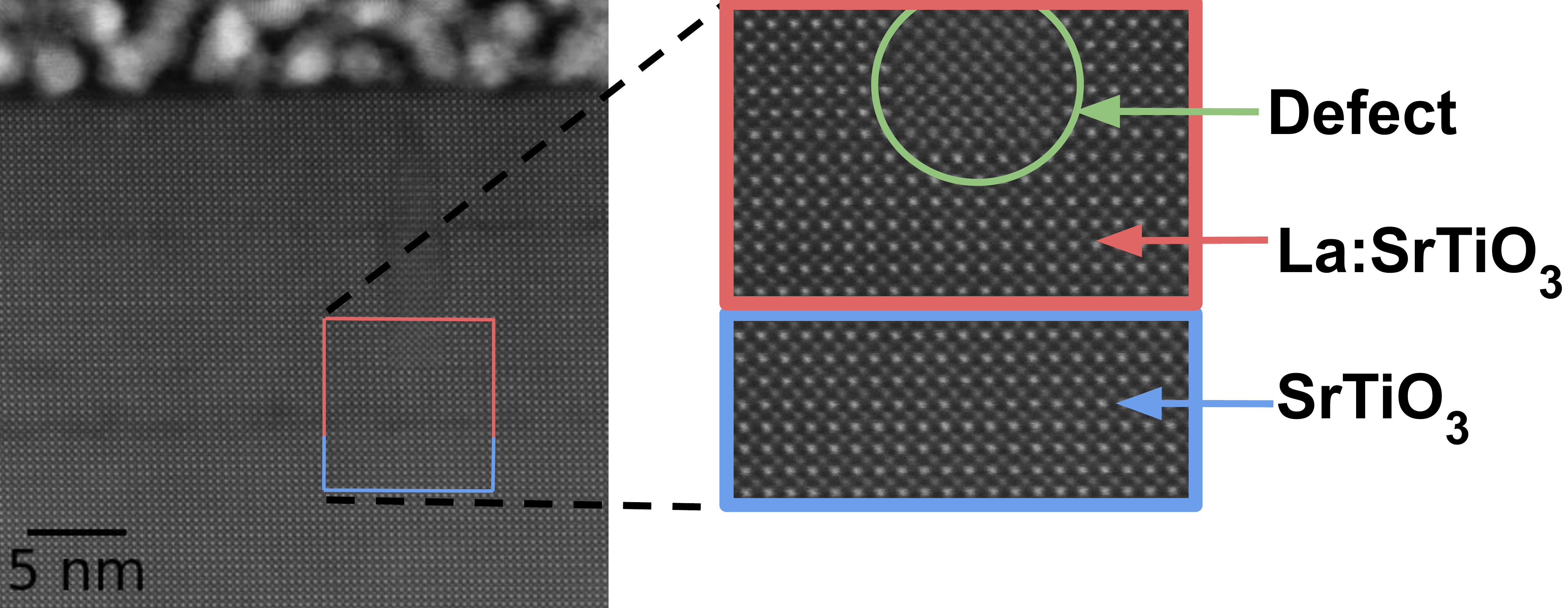}
    \caption{A defect appears in the middle.}
    \label{fig:diff_b}
  \end{subfigure}
  \caption{Two sample images of difficulty level 3. SAM fails to differentiate the component materials in both cases.}
  \label{fig:difficult}
\end{figure}

Upon closer inspection of images that are hard to segment, we discovered why SAM and SAM-I-Am were not performing well. Certain images had very narrow ground truth masks that made it difficult to detect given our sparse grid prompt. Others appear virtually indistinguishable. In Fig.~\ref{fig:difficult}, for instance, the La:SrTiO$_3$ and SrTiO$_3$ crystalline surfaces seem identical for the untrained eye. Moreover, in some cases, unidentified defects that appear in the middle of other crystalline surfaces happen to evade SAM's radar. 

These instances are predominantly observed in difficulty classes 2 and 3, leading to a significant reduction in performance. In light of this, we assigned weights to ground truth labels by classifying as hard and easy. The easy ones were assigned 10$\times$ weight while the hard ones were assigned 1$\times$. This weighted averaging provided a means to run a par experiment to determine how much loss was sustained due to the hard labels. The results are presented in Fig.~\ref{fig:performance}.

\begin{figure}[t]
    \centering
    \includegraphics[width=\columnwidth]{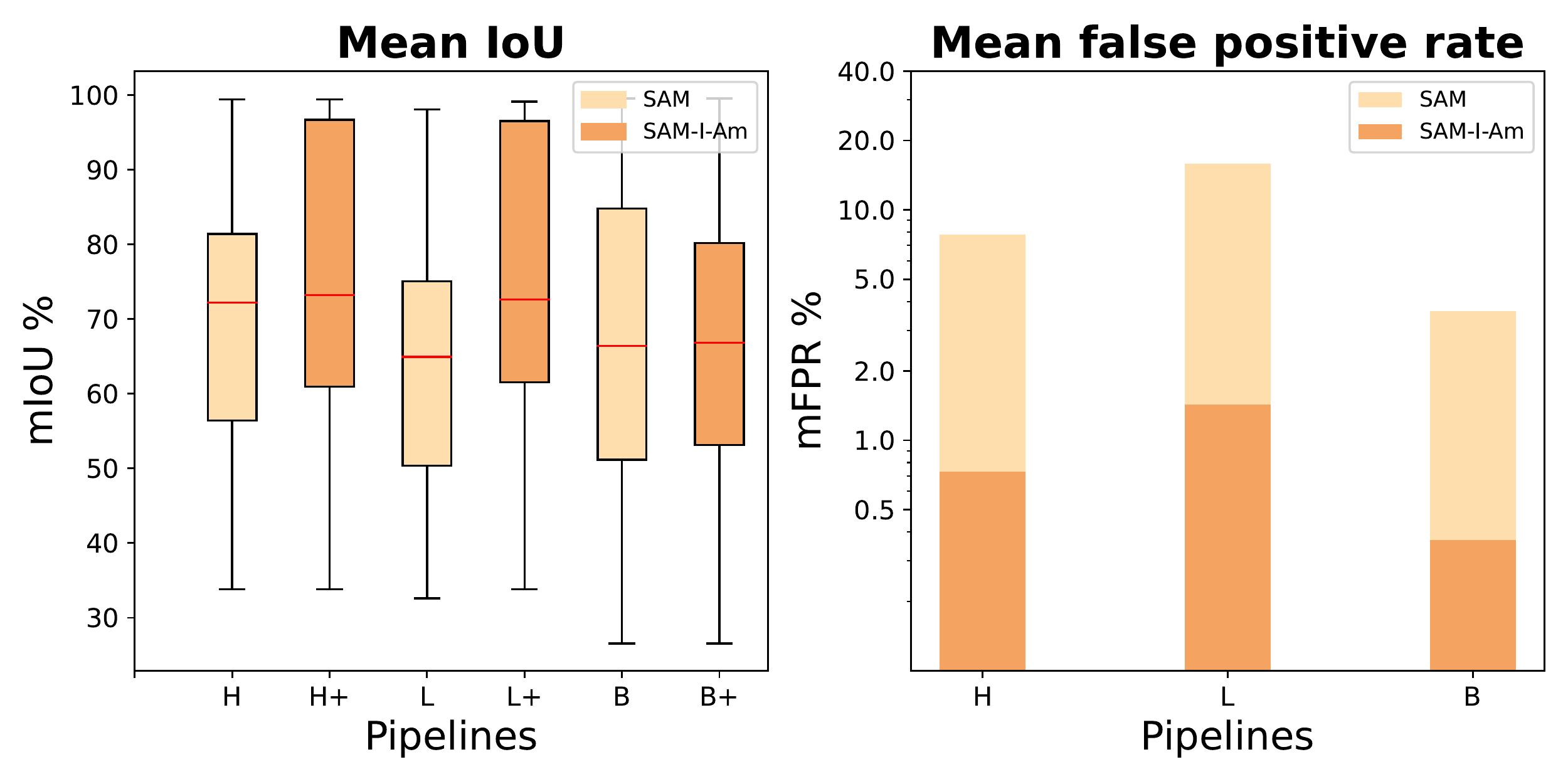}
    \caption{The semantic booster enhances performance in all SAM versions (i.e. ViT-H, ViT-L, and ViT-B).}
    \label{fig:model_v_performance}
\end{figure}


\section{Ablation study}
\label{sec:ablation}

\subsection{Varying model sizes}

The SAM ViT-B version is the smallest of the three SAM architectures, it has the advantage of segmenting faster but with less accuracy. In contrast, the larger image encoders provide improved image representation which helps produce more fine-grained masks. In this experiment, we compare the mIoU and mFPR measures of the three variants with and without the post processor engine.

Figure \ref{fig:model_v_performance} shows SAM-I-Am makes significant gains with SAM-H and SAM-L versions in terms of mIoU. On the other hand, mFPR values are reduced in all ViT versions. With the SAM-B however, there is minimal gain in mIoU. Upon closer inspection, we discovered that the SAM-B was unable to properly segment certain materials such as Pt / C. It would instead only segment bits of such materials making it impossible to correct with the SAM-I-Am booster. Consequently, both pipelines end up with comparable mIoUs. This indicates that SAM-I-Am's performance is contingent upon its inputs. In particular, SAM-I-Am achieves optimal results when its input masks effectively cover valid regions, albeit in the presence of significant mismatches and ambiguities.



 \subsection{Semantic booster with supervised classifier}


We experimented with an alternative booster that utilizes a supervised crop classifier in lieu of the unsupervised crop clustering approach in Fig.~\ref{fig:pipeline}. The classifier was trained on simulated crops of SrTiO$_3$ and Ge crystalline structures and applied as crop classifier preceding the mask-merging step. 

An interesting observation is 
a large performance discrepancy between pre-trained classifiers (i.e. pre-trained on ImageNet) vs randomly initialized classifiers. In both ResNet and ResNet-based FENet classifier, the pre-trained classifiers performed poorly after fine-tuning them on simulated data, potentially indicating of a pre-training bias. On the contrary the classifiers performed well when initialized randomly and trained on the simulated data. In the latter case, classifiers exhibited comparable performance of $\geq$80\%. 

Since we consider a majority vote for crop classification, an 80\% accuracy can make a significant impact in accurately classifying the masks. This alternative ultimately relies on the simulation cost and quality as well as the classifier's performance.

\begin{figure}[t]
    \centering
    \includegraphics[width=0.3\textwidth]{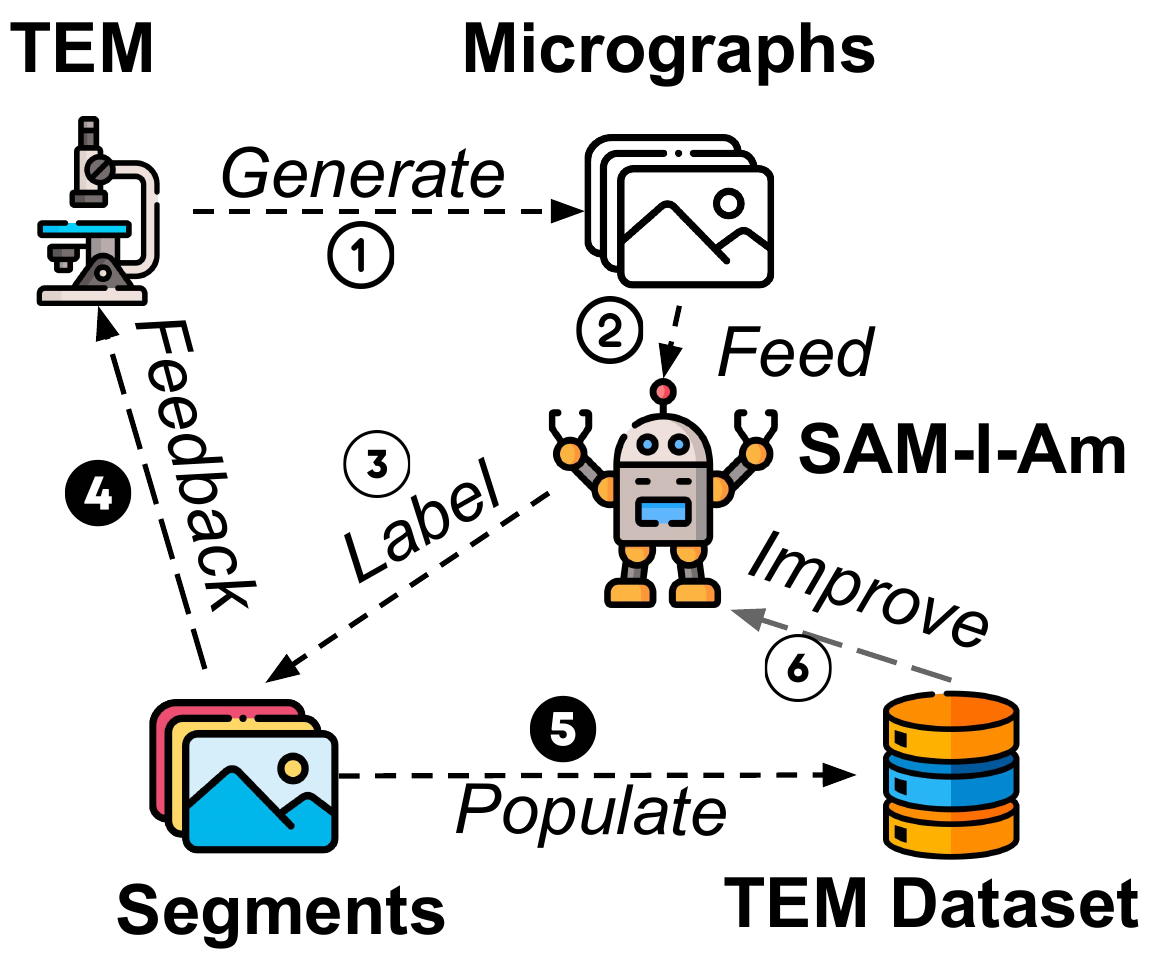}
    \caption{Injecting SAM-I-Am into the data collection loop.
    }
    \label{fig:future}
\end{figure}

\vspace{-1em}

\subsection{Future work}

Scarcity of labeled data and the combinatorial complexity of material spaces hampers progress in material discovery \cite{gomes2019artificial}. Despite significant increase in TEM microscope throughput, labeling raw data remains tedious and expensive. Among other things, we believe SAM-I-Am can facilitate the process of compiling a universal TEM dataset. In doing so, we can think of at least 2 applications. In Fig. \ref{fig:future}, arrow \ding{205} shows an application whereby SAM-I-Am helps inform the microscope about the region of interest (ROI) in real-time, an important feedback for optimizing TEM. On the other hand, arrow \ding{206} shows SAM-I-Am assisting in the compilation of a TEM dataset. Such a dataset can spur several downstream tasks including fine-tuning SAM ergo improving SAM-I-Am itself.


\section{Conclusions}
\label{sec:conclusions}

Segmentation of transmission electron microscopy (TEM) images is a challenging task due to the various characteristics of the images including intricate and often indiscernible patterns, noise, zoom levels, deformities, and so on. Moreover, the combinatorial nature of the materials and the cost of preparing labeled data impede the efficacy of deploying conventional deep learning models. In this paper, we first define the problem as a promptable segmentation task called \textit{microstructure segmentation}. Next, we deploy the Segment Anything Model (SAM) pipeline as a zero-shot solution.

Despite SAM's remarkable zero-shot performance in identifying contiguous regions, we observe it suffers from mask ambiguity and superfluous mask generation. To address these limitations, we implement a semantic booster that considers various geometric and textural information of the predicted masks in order to perform mask removal and mask merging operations. The booster significantly improves the zero-shot performance of the downstream task. Consequently, we gain a +21.35\%, +12.6\%, +5.27\% in mean IoU, and a -9.91\%, -18.42\%, -4.06\% drop in mean false positive masks across images of three difficulty classes over vanilla SAM (ViT-L). Our contributions open up new horizons for TEM research and showcase the adaptability of promptable foundation models for specialized domains. 

\bibliographystyle{icml2024}

\bibliography{refs}




\end{document}